# A tropical « NAT-like » belt observed from space

Chepfer, H.[1] and Noel, V.[2]

[1] Laboratoire de Meteorologie Dynamique / IPSL, UPMC.

[2] Laboratoire de Meterologie Dynamique / IPSL, CNRS.

Corresponding author:   H. Chepfer,

Laboratoire de Meterologie Dynamique

Ecole Polytechnique

91128 Palaiseau

France

helene.chepfer@lmd.polytechnique.fr

**Abstract.**

The optical properties of cold tropical tropopause clouds are examined on a global scale, using two years of space-borne lidar observations from CALIPSO (June 2006 - May 2008). The linear depolarization ratio, color ratio and backscatter signal are analyzed in tropical clouds colder than 200 K in a way similar to recent studies of Polar Stratospheric Clouds (PSCs). It is found that the three categories of particles encountered in PSC (Ice, Sulfate Ternary Solutions or STS, and Nitric Acid Trihydrate or NAT) do also occur in tropical cold cloud layers. Particles with optical properties similar to NAT are few, but they cover the tropical belt and represent about 20% of cold cloud tropical particles all year long. The optical behavior of these particles requires them to be very small, non-spherical, optically thin, and persistent in the TTL at temperatures colder than 200 K; NAT particles and very small ice crystals meet these criteria and are right now the best candidates to explain the presented observations.

## 1. Introduction

As NAT particles play an important role in the vertical nitric acid redistribution (Krämer et al. 2008) and in the regulation of nitric acid and water vapour (Krämer et al. 2006), with a possible impact on the radiative budget at the top of the troposphere (Gao et al. 2004), it is a key step to get a global quantitative view of NAT-containing particles in the tropics.

The possibility of observing NAT particles in the Tropical Tropopause Layer (TTL) with space-borne lidar was first proposed by Omar and Gardner (2001) and Hervig and McHugh (2002), based on the analysis of a limited number of lidar profiles without polarization capability. Their interpretation of this dataset was controversial (Jensen and Drdla, 2002) because the amount of NAT that can form in TTL conditions is so small that it seemed hardly detectable from space. Since then, several in situ observations have confirmed the existence of NAT in very small amount in the TTL (Popp et al. 2004, Popp et al. 2007, Voigt et al. 2008) as predicted theoretically by Hamill and Fiocco (1988). TTL NAT particles were also observed from a ground-based polarization lidar (Immler et al. 2007).

For NAT detection purposes, the new CALIOP / CALIPSO lidar (Winker et al. 2007) presents two advances compared to previous lidars in space: it measures the linear depolarization ratio and has collected a statistically significant dataset (more than 2 years) in the tropics and polar regions where NAT are known to occur during winter in PSCs (Voigt et al. 2000, Höpfner et al. 2006).

Here we analyze CALIPSO observations in a similar way in Austral PSCs and cold TTL clouds (T < 200 K) to evaluate how frequently NAT-like particles occur therein on a global scale. The detection of NAT-like particles uses two constraints (lidar backscatter intensity and linear depolarization ratio) following a classification method that has proven to be effective to discriminate ice, NAT and STS particles in PSC from ground-based (e.g. Stein et al. 1999, Massoli et al. 2006), airborne (Browell et al. 1990) and space-borne lidar observations (e.g. von Savigny et al. 2005, Pitts et al. 2007).

We first detect optically thin clouds colder than 200 K (Sect.2), then we study their optical properties in tropical and polar regions (Sect.3), before focusing on the NAT-like particles spatial distribution in the tropics (Sect. 4). Results are discussed and conclusions drawn in Sect. 5.

## 2. TTL clouds and PSCs colder than 200 K

TTL clouds and PSCs were detected for all CALIPSO night-time observations collected between June 2006 and May 2008. CALIPSO follows a nearly sun-synchronous orbit which crosses the equator at 01:30 Local Solar Time (night time orbits), two successive orbits are closer in space at polar latitudes than in the tropics. The altitudes of each cloud layer top ($z_{top}$) and base ($z_{base}$) are derived first, then clouds are selected based on mid-layer temperatures.

### 2.1. Cloud detection

CALIOP level 1 data are a succession of vertical profiles of altitude-dependent backscattered signal, each a vertical stack of individual lidar data points. Each CALIOP point describes the optical properties of a slice of the atmosphere at a given altitude; near the tropopause, such a slice is 60 meters thick.

For TTL clouds from 30°S to 30°N, $z_{top}$ and $z_{base}$ were derived from the level 1B product Attenuated Total Backscatter at 532 nm ($ATB_{532}$). $ATB_{532}$ profiles were averaged horizontally over 5 km to improve the signal-to-noise ratio (the initial horizontal resolution is 330 m along track and 75 m across track), then $z_{top}$ and $z_{base}$ were obtained by thresholding the signal at 8 10$^{-4}$ km$^{-1}$ sr$^{-1}$ above the molecular background profile (as in Noel et al. 2008). Cloud layers were then filtered to keep only those at least 180 m vertically and 15 km horizontally. This method closely reproduces cloud detections from the CALIOP level 2 cloud product, but increases the number of optically thin clouds detected. Mid-layer temperature was derived from GMAO GEOS-5 temperatures included in the CALIOP level 1B product.

For PSCs, $z_{top}$, $z_{base}$ and the mid-layer temperature were directly read from the CALIOP level 2 aerosol product at 5 km horizontal resolution (the vertical resolution is 60 m for altitudes below 20.2 km and 180 m above). Only observations over the south polar region (60°S to 82°S) are used here since the largest population of PSCs is found there, and CALIOP data has already proven itself useful in their study, e.g. Pitts et al. (2007) and Noel et al. (2008).

From now on, only clouds with mid-layer temperature colder than 200 K are considered ; in the TTL these clouds are detected between 14 and 20 km with a peak at 16 km.

2.2. Occurrence of TTL clouds colder than 200K

Lidar data points between $z_{base}$ and $z_{top}$ are called "cloud points" from here onwards; $N_{cpts}$ is the number of clouds points in a profile and $<N_{cpts}>$ its seasonal average in lat-lon cells. $<N_{cpts}>$ for January-February-March (JFM) and July-August-September (JAS) shows cold clouds are mostly located over areas of strong convection (Fig. 1) - South America, Africa and the Western Pacific in JFM and higher latitudes in JAS. $<N_{cpts}>$ quickly drops for latitudes poleward of 20° and is close to zero for midlatitudes, which suggests the cloud detection scheme is effective and unaffected by instrumental or algorithmic bias. $<N_{cpts}>$ is maximum in January (month of maximum convection) and minimum in August (down to ~25-30% of January values), with a near-linear evolution in between. The zonal distribution follows the Inter-Tropical Convergence Zone (ITCZ), peaking at 8°S in JFM (extending from 20°S to 10°N) and at 12°N in JAS (10°S to 25°N); this is consistent with previous space-borne lidar cloud detections near the TTL using the Geoscience Laser Altimeter System (Dessler et al. 2006). More cold clouds are observed in JFM than JAS following the tropopause temperature variation.

## 3. Particle optical properties in PSCs and TTL cold clouds

Distributions of scattering ratio $R_{532}$ (ratio between $ATB_{532}$ and molecular backscatter) and particulate depolarization ratio $\delta$ (ratio between perpendicular and parallel backscatter) are shown in Fig. 2 for cloud points in PSC layers (first row) and cold TTL clouds (second row) accumulated over the two-years CALIOP dataset. The tilting of the CALIPSO platform from 0.3° away from nadir to 3° in November 2007 does not affect $R_{532}$ and $\delta$ in the cloud layers under study. This is expected since tilting should only affect the retrieval of optical properties in planar crystals (Hu et al. 2007), which exist at warmer temperatures.

The PSC distribution for T < 200 K (Fig. 2, top left) follows a now well-known pattern that highlights the variety of particle optical properties found in these clouds. Most interpretations of this distribution agree on its dominant features (e.g. Browell et al. 1998, Stein et al. 1999, Dornbrack et al 2002, Massoli et al. 2006): ice crystals PSCs (Type II) appear as the top-right sharp edge ($\delta$ > 0.3 and $1-1/R_{532}$ > 0.7), small STS liquid particles are spread at the bottom ($\delta$ < 0.05), and the rest of the distribution includes various types of $HNO_3$-containing particles (NAT, NAT-Rock, enhanced NAT, etc). Restricting the temperature range between 195 and 200 K (Fig. 2, top right) effectively removes ice-based particles from the distribution and keeps NAT and STS-based ones, in line with the stratospheric ice nucleation temperature threshold (~188 K, Alfred et al. 2007) relevant to $HNO_3$ and water vapour concentrations encountered in the polar winter.

For TTL clouds (Fig. 2, bottom row), patterns are qualitatively different: the distribution shows two clear maxima, one roughly circular-shaped located at ($0.3 < \delta < 0.7$ and $1-1/R_{532} > 0.7$) and a smaller one more rectangular for $\delta < 0.05$. These two maxima fit with the criteria used in polar regions: the top-right maximum can be attributed to large ice-based crystals (high depolarization and scattering ratios), the bottom maximum to small liquid sulfate-based aerosols ($\delta < 0.05$) common in the upper troposphere (Chen et al. 1998), and the rest of the distribution to particles similar to NAT in their optical behavior (called "NAT-like" in the rest of this article). The distribution does not change much with temperature (Fig. 2, bottom row) because the ice nucleation temperature threshold in the TTL is much warmer than 195 K, the TTL being richer in water vapour than polar stratosphere. Ice-based particles are more numerous because the lower nitric acid and larger water vapour amounts should limit the formation of NAT and STS in the TTL. Nevertheless, when accumulated over two years, the number of TTL cloud points with NAT-like properties is significant (Fig. 2d).

In the TTL, the color ratio $\chi$ observed for the three particle categories separate very cleanly (Fig. 3), even though it was not used in the classification: $\chi \sim 1$ for large ice-based crystals (similar to observations in cirrus clouds, Tao et al. 2008), $\chi < 0.6$ for small sulfate-based aerosols (Liu et al. 2002). NAT-like particles produce $0.2 < \chi < 0.8$ which indicates intermediate sizes. This clean separation is a strong confirmation that each category describe a different particle type.

**4. Spatio-temporal distribution of NAT-like particles in the tropics**

NAT-like cloud points appear at altitudes between 15 and 20 km below 200 K all along the tropical belt in the deep convection regions. During JFM, the percentage of cloud points identified as NAT-like (Fig. 4) is 20% in average. NAT-like percentage remains constant all over the year, but the number of cold cloud points is so small in JAS (Fig. 1b) that NAT-like points are almost negligible during this season. This percentage and spatial distribution are consistent with results from Chepfer et al. (2007).

Moreover, both the NAT-like spatial distribution (Fig. 4) and the values of color ratio (Fig. 3b) are in extremely good agreement with the tropical tropopause NAT belt suggested by Voigt et al. 2008 (Fig. 4 and 7) - i.e. the particle size measured in situ and the NAT-like geographical distribution predicted by global chemistry model.

## 5. Discussion and conclusion

Two years of space-borne lidar observations have been analyzed statistically. The dataset corresponding to TTL cold clouds was split into three categories associated with clear distinct distributions of depolarization, scattering and color ratios. Those distributions fit criteria commonly used to discriminate optical properties of ice, sulfate-based and NAT-containing particles in PSCs. Distributions in cold TTL clouds and in PSCs differ in two points: the relative weight of each category (ice vs sulfate-based vs NAT-like) and their temperature dependence around 195 K. Both differences are consistent with expectations that ice and sulfate-based particles are more numerous than NAT-like particles in the TTL, and that the ice nucleation threshold of about 195 K in PSCs is warmer in the tropics.

Those particles are a minority within TTL cold clouds: 20% of cloud points below 200 K. However, they are found in the entire tropical belt, along the ITCZ where the TTL temperature is coldest.

The nature of these particles cannot be strictly asserted yet because optical properties measurements can have several physically realistic interpretations, however some possibilities can be ruled out: they are not spherical aerosols because of the depolarization ratio; they are not large ice crystals (because of both depolarization and color ratios), they are likely not dust aerosols because of their spatial distribution in the upper troposphere at all longitudes. Very small (diameter < 2 µm) non-spherical ice crystals would likely satisfy the optical properties observed, but this hypothesis is not fully satisfactory because theoretical considerations suggest these particles should not be stable (due to evaporation or sedimentation); nevertheless this possibility cannot be excluded yet as the stability of particles within the very slow non-convective vertical motions of the TTL is not well known. More "exotic" particles could also be involved, such as organic droplets (Murphy 2005, Murray et al. 2008), but their optical properties as they would be seen by a lidar are unknown yet.

Keeping in mind the recent literature showing the presence of NAT-containing particles in TTL clouds, NAT particles are today the best candidates to explain the present measurements, thus suggesting the existence of a tropical NAT belt. If these are not NAT but very small ice particles, the question on how do the ice crystals persist with such small equivalent diameter remains open. Either way, the particles in question will most certainly play a role in the regulation of the water vapour and the ozone budget at the TTL, two key components of the climate system.

***Acknowledgments.***

We thank the two anonymous reviewers for their very constructive comments. We thank the ICARE thematic center and the Climserv computing facility for providing access to the CALIOP data and to their hardware resources.

**Figure 1:** Number of cloud points per profile <N$_{cpts}$> with mid-layer temperature colder than 200 K averaged in 2.5°x2.5° cells for the (a) January-to-March (JFM) (b) July-to-September (JAS) 2007 and 2008 periods. Cells with less than 1000 cloud layers are not shown. Clouds are detected between 14 and 20 km of altitude with a peak at 16 km.

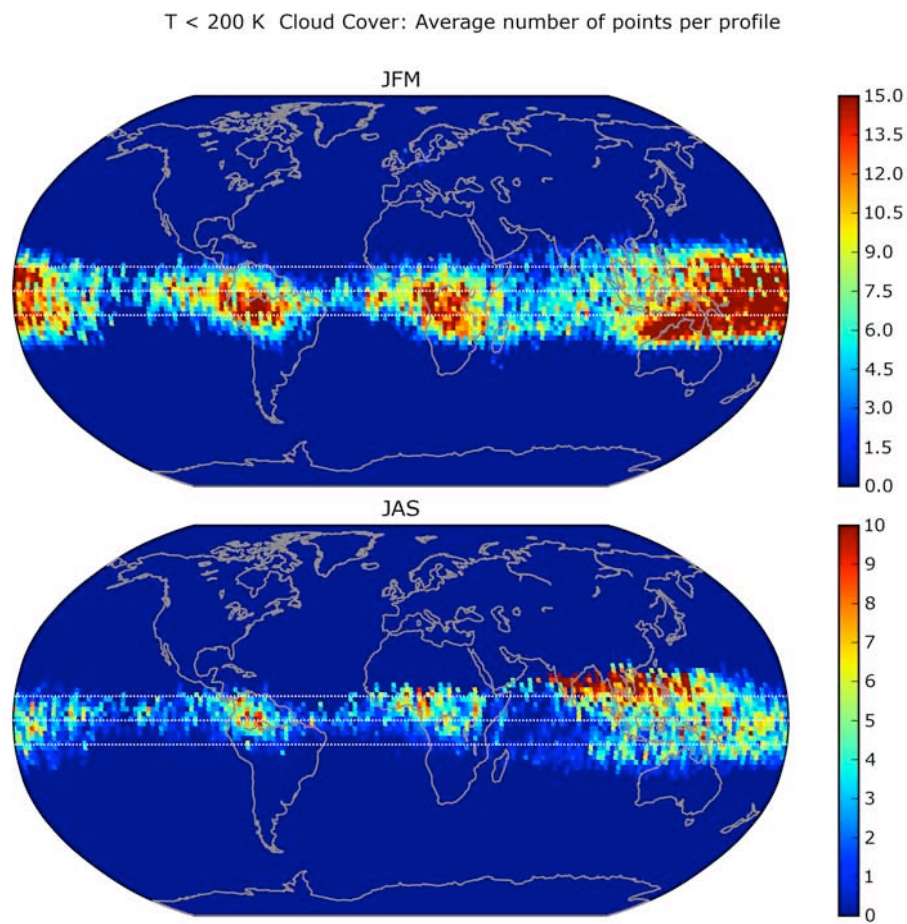

**Figure 2**: Distributions of scattering ratio $R_{532}$ (ratio between $ATB_{532}$ and molecular backscatter, shown as $1-1/R_{532}$ following Adriani et al. 2004) and depolarization ratio δ (ratio between perpendicular and parallel backscatter) for PSCs (upper row) and TTL clouds (bottom row) with mid-layer temperature colder than 200 K (left column) and in the 195-200 K range (right column) using all in-cloud observations from June 2006 to May 2008. The color scale (different in each plot) indicates the number of cloud points producing a given (depolarization, backscattering ratio) set. Small values of $1-1/R_{532}$ (x-axis) correspond to optically thin clouds.

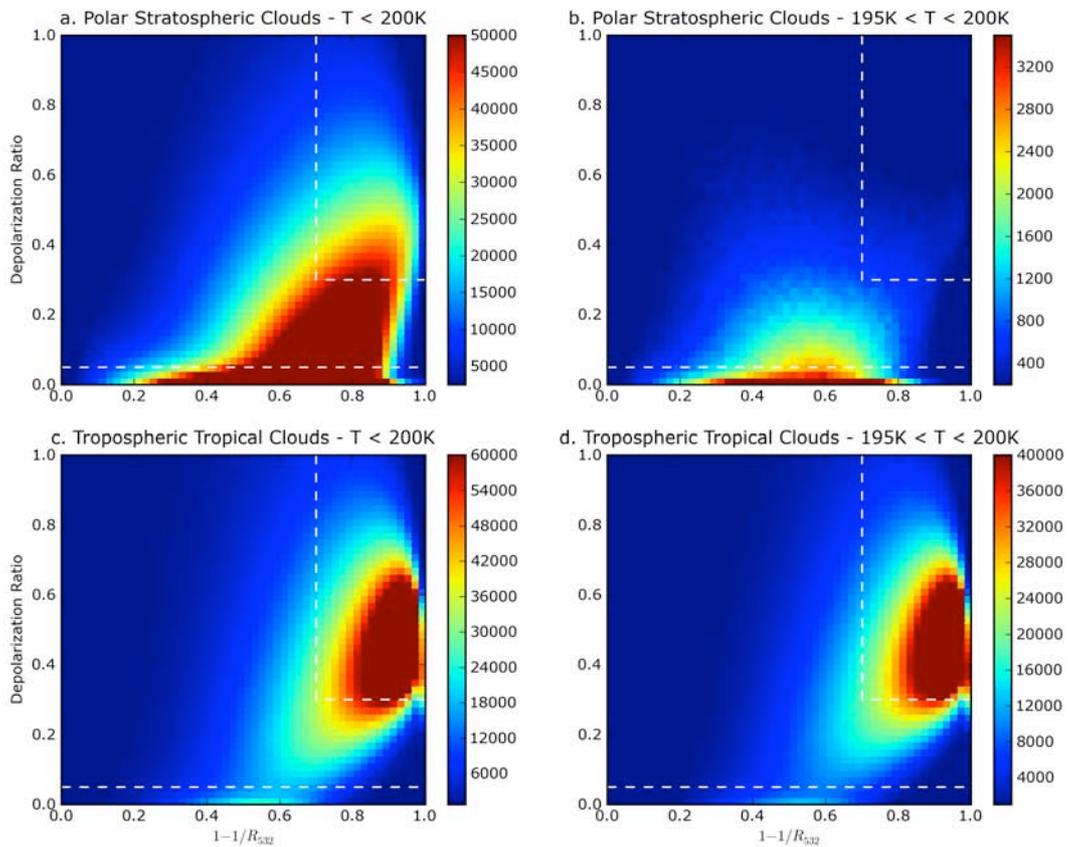

**Figure 3**: Distributions of depolarization δ and color ratio χ for TTL cold cloud points dominated by ice-based (top), sulfate-based (middle) and NAT-like (bottom) particles. χ is defined as the ratio of 1064 and 532 nm backscatter lidar signals and deviates from unity as particle size gets closer to lidar wavelength. The color scale (different in each plot) indicates the number of cloud points producing a given (depolarization, color ratio) set.

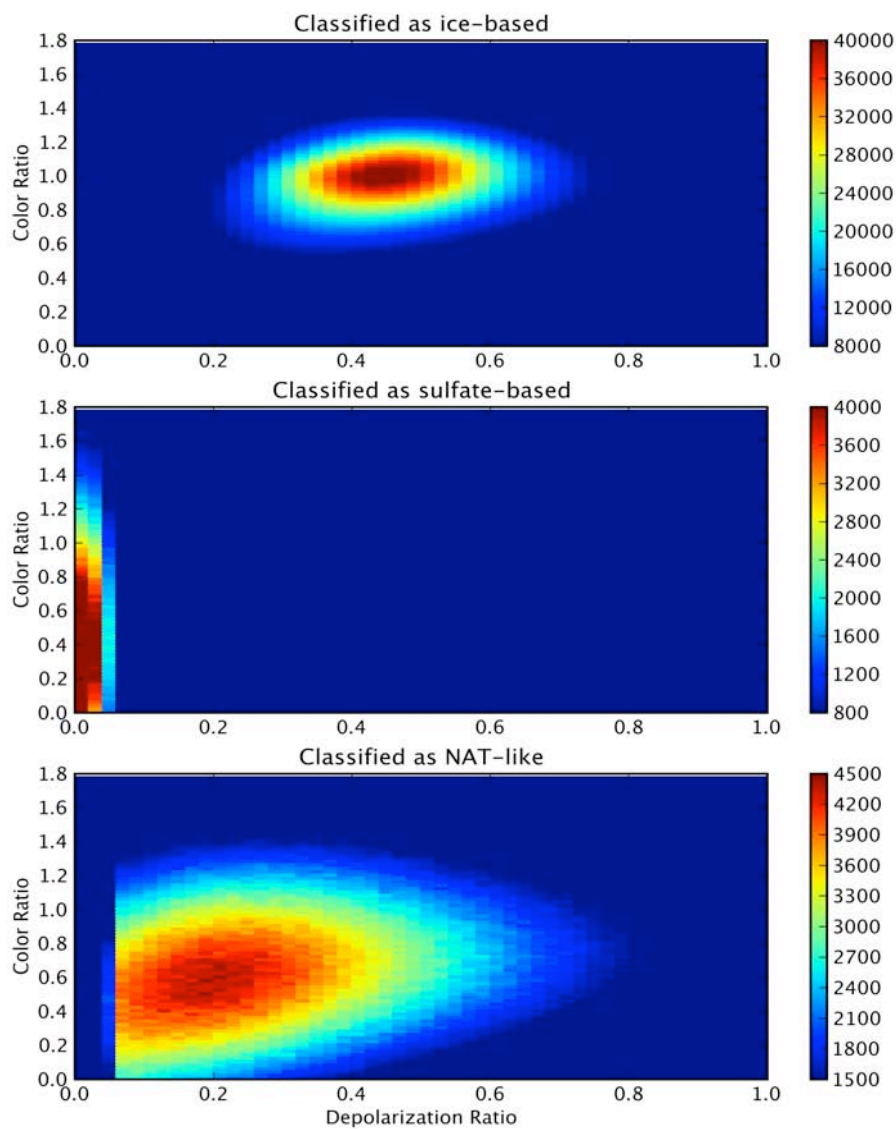

**Figure 4** : Percentage of cloud points identified as NAT-like within the TTL cold clouds shown in Fig. 2 in 2.5°x2.5° cells for JFM 2007 and 2008 periods. Cells with less than 1000 cloud layers are not shown.

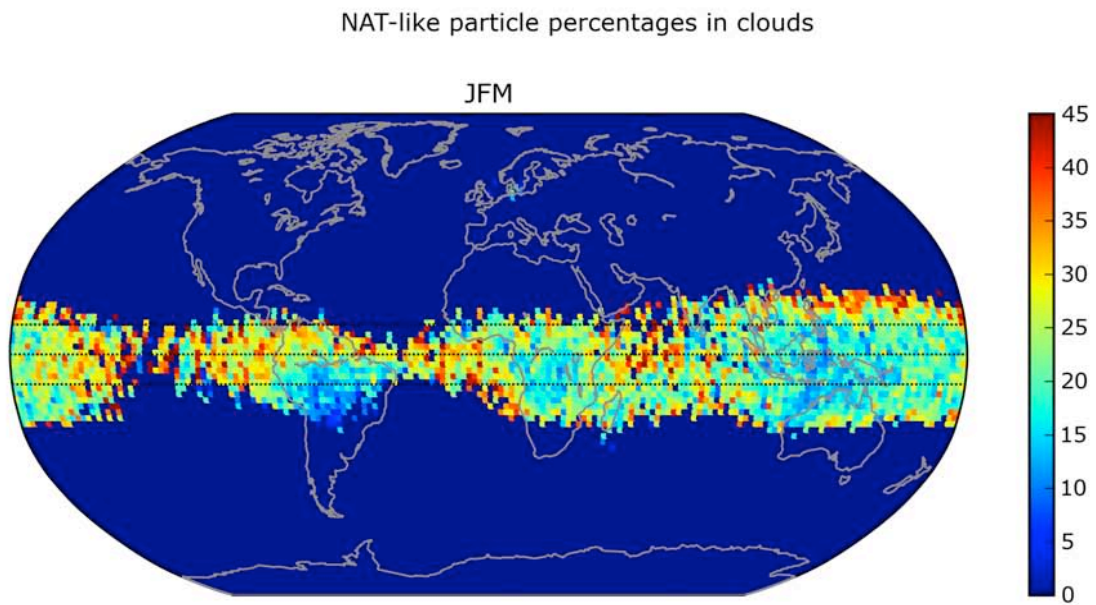